\documentclass{jetpl}
\twocolumn

\usepackage[T1]{fontenc}
\usepackage[cp1251]{inputenc}
\usepackage[russian]{babel}

\usepackage{amsmath,amssymb,epsfig,epsf}

\title{Ginzburg -- Landau expansion and the upper critical field in
disordered attractive Hubbard model.
}

\rtitle{Ginzburg -- Landau expansion and the upper critical field in
disordered attractive Hubbard model.
}

\sodtitle{Ginzburg -- Landau expansion and the upper critical field in
disordered attractive Hubbard model.
}

\author{N.\ A.\ Kuleeva\thanks{E-mail: strigina@iep.uran.ru},
E.\ Z.\ Kuchinskii\thanks{E-mail: kuchinsk@iep.uran.ru},
M.\ V.\ Sadovskii\thanks{E-mail: sadovski@iep.uran.ru}}

\rauthor{N.\ A.\ Kuleeva, E.\ Z.\ Kuchinskii, M.\ V.\ Sadovskii}

\sodauthor{N.\ A.\ Kuleeva, E.\ Z.\ Kuchinskii, M.\ V.\ Sadovskii}

\sodauthor{Н.\ А.\ Кулеева, Э.\ З.\ Кучинский, М.\ В.\ Садовский}

\address{Institute for Electrophysics, RAS, Ural Branch, Amundsen str. 106,
Ekaterinburg, 620016, Russia}

\abstract{We present a short review of our studies of disorder influence upon
Ginzburg -- Landau expansion coefficients in Anderson -- Hubbard model with attraction
in the framework of the generalized DMFT+$\Sigma$ approximation. A wide range
of attractive potentials $U$ is considered -- from weak coupling limit, where
superconductivity is described by BCS model, to the limit of very strong
coupling, where superconducting transition is related to Bose -- Einstein
condensation (BEC) of compact Cooper pairs, which are formed at temperatures
significantly higher than the temperature of superconducting transition, as well
as the wide range of disorders -- from weak to strong, when the system is in the
vicinity of Anderson transition. For the same range of parameters we study in
detail the temperature behavior of orbital and paramagnetic upper critical
field $H_{c2}(T)$, which demonstrates the anomalies both due to the growth of
attractive potential and the effects of strong disordering.}


\begin{document}

\maketitle

\section{Introduction}

The studies of disorder influence upon superconductivity have rather long history.
In classic papers by Abrikosov and Gor'kov \cite{AG_impr,AG_imp,Gor_GL,AG_mimp} 
the weak disorder limit ($p_Fl\gg 1$, where $p_F$ is Fermi momentum and
$l$ is the mean free path) was considered for the case of weak coupling 
superconductivity, which is well described by BCS theory.
The notorious ``Anderson theorem'' on the critical temperature $T_c$ of superconductors
with ``normal'' (nonmagnetic) disorder \cite{And_th,Genn} is also related to this limit.
The generalization of the theory of ``dirty'' superconductors to the case of strong
enough disorder ($p_Fl\sim1$) (and further, up to the vicinity of Anderson transition) 
was made in Refs. \cite{SCLoc_1,SCLoc_2,SCLoc_3,SCLoc_4}, where superconductivity was also
considered in the weak coupling limit.

The problem of BCS theory generalization to the region of very strong coupling is
also analyzed for a long time. Significant progress in this direction was achieved
in a paper by Nozieres and Schmitt-Rink \cite{NS}, who proposed an effective method
to study the crossover from BCS behavior in the weak coupling limit to Bose --
Einstein condensation (BEC) in the region of strong coupling. At the same time, the
problem of superconductivity in disordered systems in the limit of strong coupling
and in the region of BCS -- BEC crossover is pretty poorly studied.

One of the simplest models to study the BCS -- BEC crossover is the Hubbard model
with attraction. The most successful approach to study Hubbard model, both to
describe the strongly correlated systems for the case of repulsive interactions
and to study BCS -- BEC crossover, is the dynamical mean field theory (DMFT) 
\cite{pruschke,georges96,Vollh10}.
In recent years we have developed the generalized DMFT+$\Sigma$ approach to Hubbard
model \cite{JTL05,PRB05,FNT06,PRB07,HubDis,UFN12,LVK16}, which is quite convenient
for the studies the role of different external (with respect to those taken into
account by DMFT) interactions. In Ref. \cite{JETP14} we used this approach to
analyze the single -- particle properties and optical conductivity of the
Hubbard model with attraction. Further on, the DMFT+$\Sigma$ method was used by us
in Ref. \cite{JTL14,JETP15} to study disorder influence on the temperature of
superconducting transition, which was calculated within Nozieres -- Schmitt-Rink
approach.

Starting with the classic paper by Gor'kov \cite{Gor_GL} it is well known that
the Ginzburg -- Landau expansion is of fundamental importance in the theory
of ``dirty'' superconductors, allowing the effective studies of the behavior of
different physical properties dependencies close to critical temperature on
disorder \cite{Genn}. The generalization of this theory to the region of strong
disorder (up to Anderson metal -- insulator transition) was also based on
microscopic derivation of the coefficients of this expansion
\cite{SCLoc_1,SCLoc_2,SCLoc_3,SCLoc_4}. However, this analysis, as noted above, 
was always done in the weak coupling limit of BCS theory.

In this paper we shall present a short review of the results obtained in our
papers \cite{JETP16,FNT16,JETP17}, devoted to microscopic derivation of the
coefficients of Ginzburg -- Landau expansion, taking into account the role of
disorder in the wide region of BCS -- BEC crossover and including the region
of strong disorder in the vicinity of Anderson transition.
We shall also review the closely related results of Refs. \cite{JETP17_Hc2,JETP18} 
on the temperature dependence of orbital and paramagnetic upper critical magnetic 
fields in the region of this crossover and for different levels of disordering.

\section{Temperature of superconducting transition}

Consider disordered nonmagnetic Hubbard model with attraction and the 
Hamiltonian:
\begin{equation}
H=-t\sum_{\langle ij\rangle \sigma }a_{i\sigma }^{\dagger }a_{j\sigma
}+\sum_{i\sigma }\epsilon _{i}n_{i\sigma }-U\sum_{i}n_{i\uparrow
}n_{i\downarrow },  
\label{And_Hubb}
\end{equation}
where $t>0$ is the transfer amplitude between the nearest neighbors, $U$ is
onsite attraction potential, $n_{i\sigma }=a_{i\sigma }^{\dagger }
a_{i\sigma }^{{\phantom{\dagger}}}$ is onsite number of electrons operator, 
$a_{i\sigma }$($a_{i\sigma }^{\dagger}$) is annihilation (creation) operator
of an electron with spin $\sigma$. Local energies $\epsilon _{i}$ are assumed
to be independent random variables on different lattice sites.
We assume the Gaussian distribution of energy levels $\epsilon _{i}$:
\begin{equation}
\mathcal{P}(\epsilon _{i})=\frac{1}{\sqrt{2\pi}W}\exp\left(
-\frac{\epsilon_{i}^2}{2W^2}
\right) .
\label{Gauss}
\end{equation}
Parameter $W$ here serves as the measure of disorder strength and the Gaussian
random field of energy levels creates ``impurity'' scattering, which is
considered within the standard approach, based upon calculations of the
averaged Green's functions \cite{AGD,Diagr}.

The generalized DMFT+$\Sigma$ approach \cite{JTL05,PRB05,FNT06,UFN12} extends the
standard dynamical mean field theory  (DMFT) \cite{pruschke,georges96,Vollh10} by
the addition of ``external'' self -- energy part (SEP) $\Sigma_{\bf p}(\varepsilon)$
(in general momentum dependent), which is due to any interaction outside DMFT
and provides an effective calculation method both for single -- particle and
two -- particle properties \cite{PRB07,HubDis}.

For an ``external'' SEP entering DMFT+$\Sigma$ loop, for the case of scattering by
disorder analyzed here, we use the simplest self -- consistent Born approximation,
neglecting the ``crossing'' diagrams for impurity scattering:
\begin{equation}
\Sigma_{imp}(\varepsilon)=W^2\sum_{\bf p}G(\varepsilon,{\bf p}),
\label{BornSigma}
\end{equation}
where $G(\varepsilon,{\bf p})$ is the full single -- electron Green's function in
DMFT+$\Sigma$ approximation.

To solve the effective single Anderson impurity problem of DMFT we used the
numerical renormalization group (NRG) \cite{NRGrev}.

In the following we consider the ``bare'' band with semielliptic density of states
(per unit cell and per single spin projection):
\begin{equation}
N_0(\varepsilon)=\frac{2}{\pi D^2}\sqrt{D^2-\varepsilon^2}
\label{DOSd3}
\end{equation}
where $D$ defines the halfwidth of conduction band, which is a good approximation for
for three -- dimensional case.
In Ref. \cite{JETP15} we have shown that in DMFT+$\Sigma$ approach for the model with
semielliptic density of states all the influence of disorder upon {\em single -- particle}
properties is reduced to the widening of band by disorder, i.e. to the substitution
$D\to D_{eff}$, where $D_{eff}$ is the effective halfwidth of the ``bare'' band in the
absence of electronic correlations ($U=0$), widened by disorder:
\begin{equation}
D_{eff}=D\sqrt{1+4\frac{W^2}{D^2}}.
\label{Deff}
\end{equation}
The ``bare'' (in the absence of $U$) density of states, ``dressed'' by disorder:
\begin{equation}
\tilde N_{0}(\xi)=\frac{2}{\pi D_{eff}^2}\sqrt{D_{eff}^2-\varepsilon^2}
\label{tildeDOS}
\end{equation}
remains semielliptic in the presence of disorder.

All calculations below were done for the case of quarter -- filled band
(number of electrons per lattice site $n$=0.5).

To consider superconductivity in a wide interval of pairing interaction $U$,
following Refs. \cite{JETP14,JETP15} we use Nozieres -- Schmitt-Rink approximation
\cite{NS}, which allows qualitatively correct (though approximately) describe the
BCS -- BEC crossover region. In this approach the critical temperature $T_c$ is 
determined \cite{JETP15} by the usual BCS -- like equation:
\begin{equation}
1=\frac{U}{2}\int_{-\infty}^{\infty}d\varepsilon \tilde N_0(\varepsilon)
\frac{th\frac{\varepsilon -\mu}{2T_c}}{\varepsilon -\mu},
\label{BCS}
\end{equation}
where the chemical potential $\mu$ for different values of $U$ and $W$ is
obtained from the standard equation for the number of electrons (band filling),
determined from the full Green's function, calculated in в DMFT+$\Sigma$ approximation.
This allows to find $T_c$  for a wide interval of the values of parameters of the theory,
including the BCS -- BEC crossover region and the limit of strong coupling, as well as
for different levels of disorder. It is the essence of interpolation scheme of
Nozieres and Schmitt-Rink --- in the weak coupling region transition temperature is
controlled by the equation for Cooper instability (\ref{BCS}), while in the strong
coupling limit it is determined as the temperature of BEC, which is controlled by the
chemical potential.
\begin{figure}
\includegraphics[clip=true,width=0.48\textwidth]{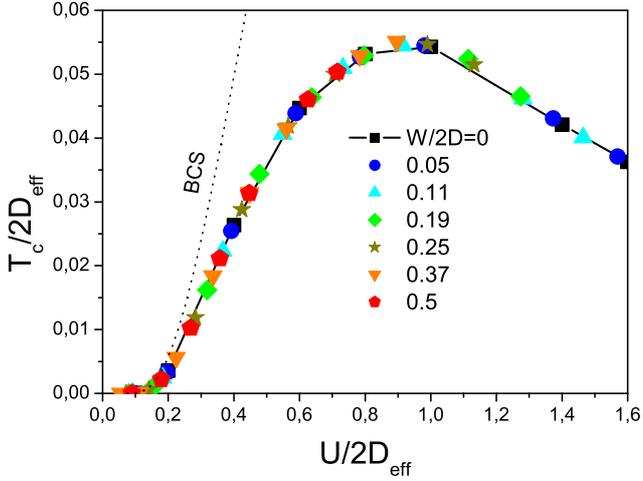}
\caption{Fig. 1. Universal dependence of the temperature of superconducting transition on
the strength of Hubbard attraction for different levels of disorder.}
\label{fig1}
\end{figure}
In Ref.  \cite{JETP15} we have shown that disorder influence on the critical temperature
$T_c$ in the model with semielliptic bare density of states is universal and is reduced just
to the change of the effective bandwidth.
In Fig. \ref{fig1}, as an illustration of this, we show the universal dependence
of critical temperature $T_c$ on Hubbard attraction for different levels of disorder,
which demonstrates the validity of the generalized Anderson theorem \cite{JTL14,JETP15}. 
In the weak coupling region the temperature of superconducting transition is well described
by BCS model (to compare in Fig.\ref{fig1} we show the dashed line corresponding to BCS
model, when $T_c$ is determined by Eq. (\ref{BCS}) with chemical potential independent of
$U$ and determined by quarter -- filling of the ``bare'' band), while in the strong coupling
region the critical temperature is mainly determined by BEC condition for Cooper pairs
and drops with the growth of $U$ as $t^2/U$, passing the maximum at $U/2D_{eff}\sim 1$. 
The review of these and some other results obtained for disordered Hubbard model in
DMFT+$\Sigma$ approximation can be found in Ref. \cite{LVK16}.

\section{Gibzburg -- Landau expansion}

Ginzburg -- Landau expansion for the difference of free energies in superconducting
and normal states can be written in a standard form \cite{Diagr}:
\begin{equation}
F_{s}-F_{n}=A|\Delta_{\bf q}|^2
+q^2 C|\Delta_{\bf q}|^2+\frac{B}{2}|\Delta_{\bf q}|^4,
\label{GL}
\end{equation}
where $\Delta_{\bf q}$ is the amplitude of the Fourier component of order parameter.
\begin{figure}
\includegraphics[clip=true,width=0.5\textwidth]{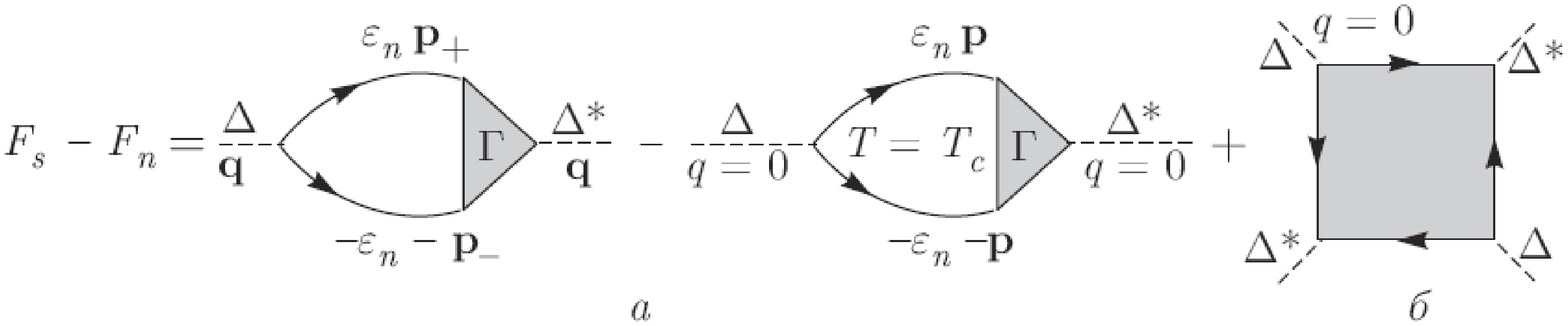}
\caption{Fig. 2. Diagrammatic form of Ginzburg -- Landau expansion.}
\label{diagGL}
\end{figure}
Expansion (\ref{GL}) is determined by diagrams of the loop expansion for free energy
in the field of fluctuations of order parameter (denoted by dashed lines) with small
wave vector  ${\bf q}$ \cite{Diagr}, shown in Fig. \ref{diagGL} \cite{Diagr}.

Within Nozieres -- Schmitt-Rink approach \cite{NS} we use weak coupling approximation
to analyze Ginzburg -- Landau coefficients, so that the loops with two and four
Cooper vertexes shown in Fig. \ref{diagGL} do not contain contributions from Hubbard
attraction and are ``dressed'' only by impurity scattering. 
However, as in the case of calculation of $T_c$, the chemical potential, which is
essentially dependent on coupling strength and in the strong coupling limit determines
the condition of Bose condensation of Cooper pairs, should be calculated within full 
DMFT+$\Sigma$ procedure. In Ref. \cite{JETP16} we have shown that in this approach 
the coefficients $A$ and $B$ are given by the following expressions:
\begin{equation}
A(T)=\frac{1}{U}-
\int_{-\infty}^{\infty}d\varepsilon \tilde N_0(\varepsilon)
\frac{th\frac{\varepsilon -\mu }{2T}}{2(\varepsilon -\mu )},
\label{A_end}
\end{equation}
\begin{equation}
B=\int_{-\infty}^{\infty}\frac{d\varepsilon}{2(\varepsilon -\mu)^3}
\left(th\frac{\varepsilon -\mu}{2T}-\frac{(\varepsilon -\mu)/2T}
{ch^2\frac{\varepsilon -\mu}{2T}}\right)
\tilde N_0(\varepsilon),
\label{B_end}
\end{equation}
For $T\to T_c$ coefficient $A(T)$ takes the following form:
\begin{equation}
A(T)\equiv \alpha(T-T_c).
\label{A2}
\end{equation} 
In BCS limit for coefficients $\alpha$ and $B$ we obtain the standard
result \cite{Diagr}:
\begin{equation}
\alpha_{BCS}=\frac{\tilde N_0(\mu)}{T_c} \qquad B_{BCS}=\frac{7\zeta(3)}
{8\pi^2 T_c^2}\tilde N_0(\mu).
\label{aB_BCS}
\end{equation} 
Thus the coefficients $A$ and $B$ are determined only by the density of states
$\tilde N_0(\varepsilon)$ widened by disorder and by the chemical potential.
For semielliptic bare density of states the dependence of these coefficients on
disorder is due only to substitution  $D\to D_{eff}$, so that in the presence of 
disorder we get the universal dependencies of $\alpha$ and $B$ (made dimensionless by
the effective bandwidth) on $U/2D_{eff}$ \cite{JETP16}. Actually the coefficients
$\alpha$ and $B$ drop fast with the growth of coupling strength $U/2D_{eff}$.
It should be noted that Eqs. (\ref{A_end}) and (\ref{B_end}) for coefficients 
$A$ and $B$ were obtained in Ref. \cite{JETP16} using exact Ward identities and
remain valid also in case of strong disorder (Anderson localization).

\begin{figure}
\includegraphics[clip=true,width=0.48\textwidth]{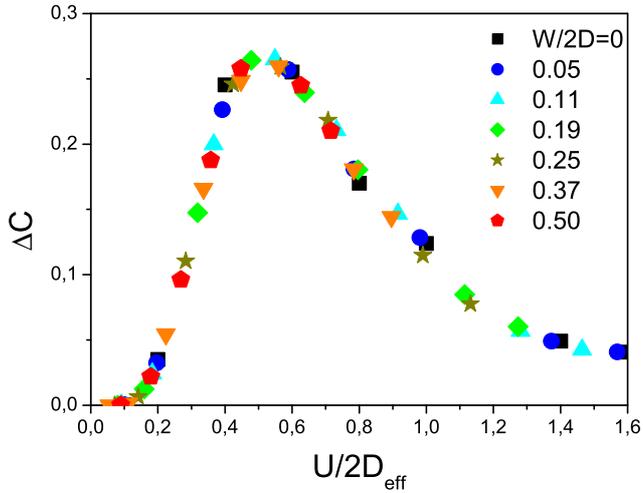}
\caption{Fig. 3. The universal dependence of specific heat discontinuity on
$U/2D_{eff}$ for different levels of disorder.}
\label{fig3}
\end{figure}

The universal dependence on disorder related to widening of the band
$D\to D_{eff}$ appears also for specific heat discontinuity at transition temperature
\cite{JETP16}, which is determined by coefficients $\alpha$ and $B$:
\begin{equation}
\Delta C\equiv C_s(T_c)-C_n(T_c)=T_c\frac{\alpha^2}{B}.
\label{Cs-Cn}
\end{equation}
This universal dependence of specific heat discontinuity on $U/2D_{eff}$ is shown in
Fig. \ref{fig3}. In BCS limit specific heat discontinuity grows with coupling
strength, while in BEC limit it drops, passing through a maximum at $U/2D_{eff}\approx$ 0.55.
This behavior of specific heat discontinuity is determined mainly by the behavior
of $T_c$ (cf. Fig.\ref{fig1}), while the ratio $\frac{\alpha^2}{B}$ in Eq. (\ref{Cs-Cn}) 
smoothly depends on the coupling strength.  

Now we shall follow Refs. \cite{FNT16,JETP17} to analyze the coefficient $C$.
From diagrammatic representation of Ginzburg -- Landau expansion, shown 
in Fig.\ref{diagGL}, it is clear that $C$ is determined as a coefficient before $q^2$ 
in Cooper -- like two -- particle loop (first term in Fig. \ref{diagGL}).
Thus we obtain the following expression:
\begin{equation}
C=-T\lim_{q \to 0}\sum_{n, \bf p, \bf {p'}} 
\frac{\Psi_{\bf p\bf {p'}}( \varepsilon_n,{\bf q})-\Psi_{\bf p\bf {p'}}
(\varepsilon_n,0)}{q^2},
\label{C1}
\end{equation}
where $\Psi_{\bf p,\bf {p'}}( \varepsilon_n,{\bf q})$ is two -- particle Green's function
in Cooper channel, ``dressed'' (in Nozieres -- Schmitt-Rink approximation) only by
impurity scattering. 

In BCS limit and in the absence of disorder the coefficient $C$ takes the following
form \cite{Diagr}:
\begin{equation}
C_{BCS}=\frac{7\zeta(3)}{16\pi^2 T_c^2}N_0(\mu)\frac{v_F^2}{d},
\label{C_BCS}
\end{equation}
where $v_F$ is velocity at the Fermi surface, $d$ is space dimensionality.
Disorder influence on coefficient $C$ is not reduced only to the substitution 
$N_0\to\tilde N_0$, so that in the presence of disorder, in contrast to
coefficients $\alpha$ and $B$ (cf. (\ref{aB_BCS})),  even in BCS limit we 
can not obtain any compact expression for $C$ similar to Eq. (\ref{C_BCS}), 

After rather cumbersome analysis \cite{FNT16,JETP17} we get the following general
expression for the coefficient $C$:
\begin{eqnarray}
C=-\frac{1}{8\pi}\int_{-\infty}^{\infty}d\varepsilon
\frac{th\frac{\varepsilon}{2T}}{\varepsilon}Im
\left(
\frac{iD(2\varepsilon)\sum_{\bf {p}}\Delta G_{\bf p}(\varepsilon)}{\varepsilon
+i\delta}
\right)=
\nonumber\\
=-\frac{1}{8\pi}\int_{-\infty}^{\infty}d\varepsilon
\frac{th\frac{\varepsilon}{2T}}{\varepsilon ^2}
Re(D(2\varepsilon)\sum_{\bf {p}}\Delta G_{\bf p}(\varepsilon))-\nonumber\\
-\frac{1}{16T}Im(D(0)\sum_{\bf {p}}\Delta G_{\bf p}(0)), 
\label{C_loc}
\end{eqnarray}
where $\Delta G_{\bf p}(\varepsilon)=G^{R}(\varepsilon,{\bf p})-
G^{A}(-\varepsilon,{\bf p})$ and  $D(\omega)$ is the frequency dependent generalized
diffusion coefficient \cite{Diagr,VW,VW2,WV,MS,MS86,KS,VW92}, which is determined within
generalization of the self -- consistent theory of localization by the following
self -- consistency equation \cite{HubDis}:
\begin{eqnarray}
D(\omega)=i\frac{<v>^2}{d}
\Biggl(\omega-\Delta\Sigma_{imp}^{RA}(\omega)+\nonumber\\
+W^4\sum_{\bf p}
(\Delta G_{\bf p}(\varepsilon))^2\sum_{\bf {q}}\frac{1}{\omega+iD(\omega)q^2}
\Biggr)^{-1},
\label{l9}
\end{eqnarray}
where $\omega=2\varepsilon$, 
$\Delta\Sigma_{imp}^{RA}(\omega)=\Sigma_{imp}^{R}(\varepsilon)-
\Sigma_{imp}^{A}(-\varepsilon)$, $d$ -- space dimensionality, and velocity
$<v>$ is determined by the following expression:
\begin{equation}
<v>=\frac{\sum_{\bf {p}}|{\bf {v_p}}|\Delta G_{\bf p}(\varepsilon)}
{\sum_{\bf {p}}\Delta G_{\bf p}(\varepsilon)}; 
{\bf {v_p}}=\frac{\partial\varepsilon (\bf p)}{\partial\bf p}.
\label{l7}
\end{equation}
Taking into account applicability limits of diffusion approximation,
summation over $q$ in Eq. (\ref{l9}) should be limited by
\cite{Diagr,MS86}:
\begin{equation}
q<k_0=Min \{l^{-1},p_F\},
\label{cutoff}
\end{equation}
where $l$ is the mean -- free path due to elastic scattering by disorder,
$p_F$  is Fermi momentum.

Thus we obtain an interpolation scheme to determine the coefficient $C$,
which in the weak disorder limit reproduces the results of ``ladder'' 
approximation, while in the strong disorder limit it takes into account the
effects of Anderson localization (in the framework of self -- consistent
theory of localization).

It was shown \cite{HubDis,UFN12} that in DMFT+$\Sigma$ approximation for
Anderson -- Hubbard model the critical disorder for Anderson metal -- insulator
transition $W/2D=$0.37 (for the choice of cutoff as in Eq. (\ref{cutoff})), so that
in this approximation it does not depend on the value of Hubbard interaction $U$.
The approach developed above allows determination of coefficient $C$ including the
region of Anderson insulator with disorder $W/2D>$0.37.

\section{Physical properties close to the temperature of superconducting
transition}

The coherence length at a given temperature $\xi(T)$ determines the characteristic
scale of inhomogeneities of superconducting order parameter:
\begin{equation}
\xi^2(T)=-\frac{C}{A}.
\label{xi2}
\end{equation}
From Eq. (\ref{A2}) we have:
$A=\alpha(T-T_c)$, то
\begin{equation}
\xi(T)=\frac{\xi}{\sqrt{1-T/T_c}},
\label{xiT}
\end{equation}
where we have introduce the coherence length of a superconductor as:
\begin{equation}
\xi=\sqrt{\frac{C}{\alpha T_c}},
\label{xi0}
\end{equation}
which in the weak coupling limit and in the absence of disorder has the
standard form \cite{Diagr}:
\begin{equation}
\xi_{BCS}=\sqrt{\frac{C_{BCS}}{\alpha_{BCS} T_c}}=\sqrt{\frac{7\zeta(3)}
{16\pi^2 d}}\frac{v_F}{T_c}.
\label{xi_BCS}
\end{equation}
The penetration depth of magnetic field into superconductor is defined as:
\begin{equation}
\lambda^2(T)=-\frac{c^2}{32 \pi e^2}\frac{B}{A C}.
\label{lambda2}
\end{equation}
Thus:
\begin{equation}
\lambda (T)=\frac{\lambda}{\sqrt{1-T/T_c}},
\label{lambdaT}
\end{equation}
where we have introduced:
\begin{equation}
\lambda^2=\frac{c^2}{32 \pi e^2}\frac{B}{\alpha C T_c},
\label{lambda0}
\end{equation}
which in the absence of disorder and in the weak coupling limit is:
\begin{equation}
\lambda^2_{BCS}=\frac{c^2}{32 \pi e^2}\frac{B_{BCS}}{\alpha_{BCS}
C_{BCS} T_c}=\frac{c^2}{16 \pi e^2}\frac{d}{N_0(\mu)v_F^2}.
\label{lambda_BCS}
\end{equation}
Note that  $\lambda_{BCS}$ does not depend on $T_c$, i.e. on the coupling strength,
and it can be conveniently used to normalize penetration depth
$\lambda$ (\ref{lambda0}) at arbitrary $U$ and $W$.

Close to $T_c$ the upper critical magnetic field $H_{c2}$ is defined via
Ginzburg -- Landau coefficients as:
\begin{equation}
H_{c2}=\frac{\Phi_0}{2 \pi \xi^2(T)}=-\frac{\Phi_0}{2 \pi}\frac{A}{C},
\label{Hc2}
\end{equation}
where $\Phi_0=c \pi/e$ is the magnetic flux quantum.
Then the slope of the upper critical field close to $T_c$ is given by:
\begin{equation}
\frac{dH_{c2}}{dT}= \frac{\Phi_0}{2 \pi}\frac{\alpha}{C}.
\label{dHc2}
\end{equation}

Coefficient $C$ is essentially a two -- particle entity, thus it is not universally
dependent on disorder in contrast to coefficients $A$ and $B$ and disorder influence
upon it does not reduce only to effective band widening by disorder.
Let us now discuss the main results of our calculations for this coefficient
(for more details cf. Refs. \cite{FNT16,JETP17}). Coefficient $C$ rapidly
decreases with the growth of coupling strength. Especially strong drop is
observed in the weak coupling region. Localization corrections become
important in the limit of strong enough disorder ($W/2D>0.25$).
For such disorder level localization corrections significantly  suppress
the coefficient $C$ in the weak coupling region, while in the strong coupling
region for $U/2D>1$ localization corrections in fact do not change the value of the
coefficient, even in the limit of strong disorder with  $W/2D>0.37$, when the system
becomes Anderson insulator. This is apparently due to the fact, that in the region
of strong coupling the (pseudo)gap is opened in the density of states at the Fermi
level \cite{JETP14}, so that there are no states to localize in the vicinity of
the Fermi level at all.
\begin{figure}
\includegraphics[clip=true,width=0.48\textwidth]{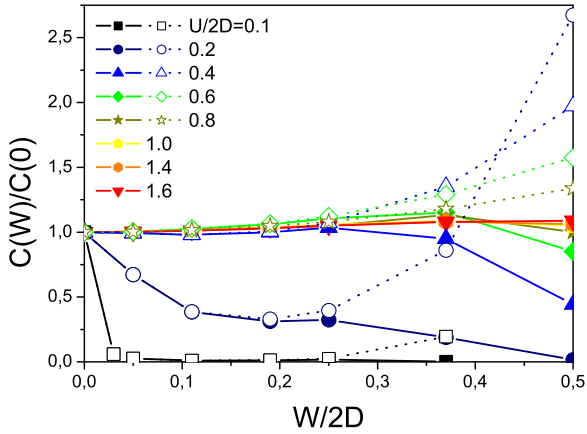}
\caption{Fig. 4. Dependence of the coefficient $C$, normalized by its value in the
absence of disorder, on disorder for different values of Hubbard attraction $U$.
Dashed line -- ``ladder'' approximation, full lines -- results obtained with account
of localization corrections.}
\label{fig4}
\end{figure}
In Fig. \ref{fig4} we show the dependencies of coefficient $C$ on disorder
strength for different values of coupling $U/2D$. In this figure (and all
that follow in this Section) the filled symbols and continuous lines correspond
to calculations taking into account localization corrections, while the empty
symbols and dashed lines correspond to ``ladder'' approximation.
In the weak coupling limit ($U/2D=0.1$) we observe fast enough drop of the
coefficient $C$ with disorder growth in the region of weak impurity scattering.
At the same time in the region of strong enough disorder in  ``ladder'' approximation
we can observe the increase of the coefficient $C$ with the growth of disorder,
which is mainly due the noticeable widening of the band by this disorder and
corresponding drop of the effective coupling strength  $U/2D_{eff}$. 
However, localization corrections which become important for strong disorder
$W/2D>0.25$, lead to suppression of $C$ while disorder grows, also in the limit
of strong impurity scattering. In the region of intermediate coupling
($U/2D=$0.4 -- 0.6) coefficient $C$ in ``ladder'' approximation is rather
insignificantly increasing with disorder growth. In BEC limit ($U/2D>1$) 
coefficient $C$ in fact is independent of impurity scattering both in the
``ladder'' approximation and with the account of localization corrections.
Localization corrections in BEC limit in fact do not change the value of
coefficient $C$ as compared to  ``ladder'' approximation.
As Ginzburg -- Landau coefficients $\alpha$ and $B$ are universally dependent
on disorder,  Anderson localization has no influence upon them at all,
and coefficient $C$, which is strongly dependent on localization correction in
the weak coupling limit, in BEC limit is in fact independent of these corrections. 
Correspondingly, the physical properties depending on coefficient $C$, are also
significantly dependent on localization corrections in the weak coupling limit,
but in fact do not feel Anderson localization in BEC limit.

\begin{figure}
\includegraphics[clip=true,width=0.48\textwidth]{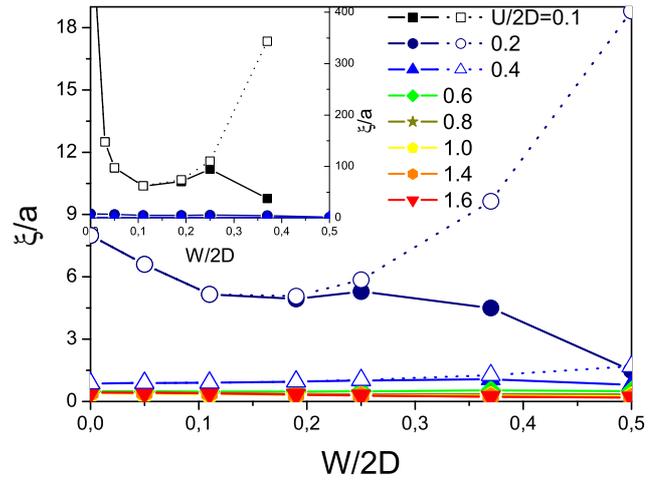}
\caption{Fig. 5. Dependence of coherence length on disorder for different values
of Hubbard attraction. Coherence length is normalized by lattice parameter $a$. 
At the insert: dependence of coherence length on disorder in the weak coupling
limit.}
\label{fig5}
\end{figure}
In Fig. \ref{fig5} we show the dependence of coherence length $\xi$ on the level
of disorder for different values of coupling strength. In the weak coupling
limit coherence length $\xi$ drops fastly with the growth of $U$ at any disorder
level, reaching the values of the order of lattice parameter $a$ in intermediate
coupling region of $U/2D \sim 0.4-0.6$. Further growth of coupling strength
only slightly changes the coherence length. In BCS limit, i.e. for the weak
coupling and weak enough impurity scattering we observe (cf. insert at 
Fig. \ref{fig5}) the standard dependence for ``dirty'' superconductors
$\xi \sim l^{1/2}$, i.e. the coherence length rapidly drops with the growth 
of disorder. However, at strong enough disorder in ladder approximation
(dashed lines) coherence length  grows with disorder, which is mainly due to
noticeable widening of the bare band and corresponding suppression of $U/2D_{eff}$. 
Localization corrections are important only for large disorder ($W/2D>0.25$) 
and lead to significant drop of coherence length in BCS limit of weak coupling
and practically does not change coherence length in BEC limit.
Taking into account localization corrections leads to noticeable drop of
coherence length as compared to ``ladder'' approximation in the limit of
strong disorder restoring the suppression of $\xi$ with the growth  of disorder
in this limit. In standard BCS model with the bare band of infinite width in the
limit weak disorder the coherence length drops with disorder
$\xi \sim l^{1/2}$, and close to Anderson transition $\xi$ drops even faster
as $\xi \sim l^{2/3}$ \cite{SCLoc_1,SCLoc_2,SCLoc_3}, in contrast to our model,
where close to Anderson transition the coherence length rather weakly depends on
disorder, which is related to a significant widening of the band by disorder.
With the growth of the coupling strength $U/2D \geq$ 0.4--0.6  the coherence
length $\xi$ becomes of the order of lattice parameter and becomes almost disorder
independent. In particular, in BEC limit of very strong coupling $U/2D=$1.4, 1.6,
the growth of disorder up to very strong values ($W/2D=0.5$) leads to a factor
two drop of coherence length, so that in the limit of strong coupling the
account of localization corrections becomes irrelevant.

\begin{figure}
\includegraphics[clip=true,width=0.48\textwidth]{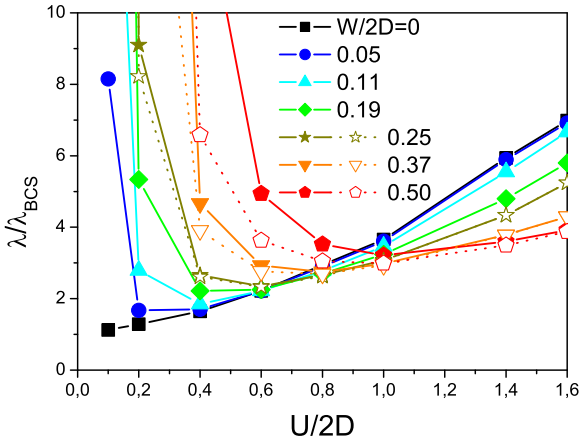}
\caption{Fig. 6. Dependence of penetration depth normalized by its BCS value
in the weak coupling limit on the strength of Hubbard attraction $U$ for different
levels of disorder.}
\label{fig6}
\end{figure}
In Fig. \ref{fig6} we show the dependence of penetration depth, normalized by its
BCS value in the absence of disorder (\ref{lambda_BCS}) on Hubbard attraction strength
$U$ for different levels of disorder. In the absence of impurity scattering
penetration depth grows with coupling strength. In the weak coupling limit, in
accordance with the usual theory of ``dirty'' superconductors, disorder leads to
a fast growth of penetration depth ($\lambda \sim l^{-1/2}$, where $l$ is the
mean free path). With increase of the coupling strength the growth of penetration
depth  with disorder slows down and in the limit of very strong coupling fro 
$U/2D=$1.4, 1.6 penetration depth even slightly decreases with the growth of disorder.
Thus, in presence of disorder we observe the drop of penetration depth with the
growth of Hubbard attraction in the region of relatively weak coupling and the
growth of $\lambda$
with $U$ in BEC strong coupling limit.
The account of localization corrections is relevant only in the limit of strong
disorder ($W/2D>0.25$) and leads significant growth of penetration depth as
compared with results of the  ``ladder'' approximation in weak coupling limit.
However, qualitatively the dependence of penetration depth on disorder does not change.
In BEC limit localization influence on penetration depth is insignificant.
Similar dependence on disorder is observed also for dimensionless Ginzburg --
Landau parameter $\kappa = \lambda / \xi$. In weak coupling limit Ginzburg --
Landau parameter rapidly grows with disorder in accordance with the theory of
``dirty'' superconductors, where $\kappa \sim l^{-1}$. With the increase of
the coupling strength the growth of Ginzburg -- Landau parameter with disorder
slows down and in the strong coupling limit of $U/2D>1$ parameter $\kappa$ is
practically disorder independent. The account of localization corrections
leads quantitatively to a noticeable increase of Ginzburg -- Landau parameter
in Anderson insulator phase ($W/2D \geq 0.37$) for the weak coupling. 
In the limit of strong coupling the account of localization is again irrelevant.

\begin{figure}
\includegraphics[clip=true,width=0.48\textwidth]{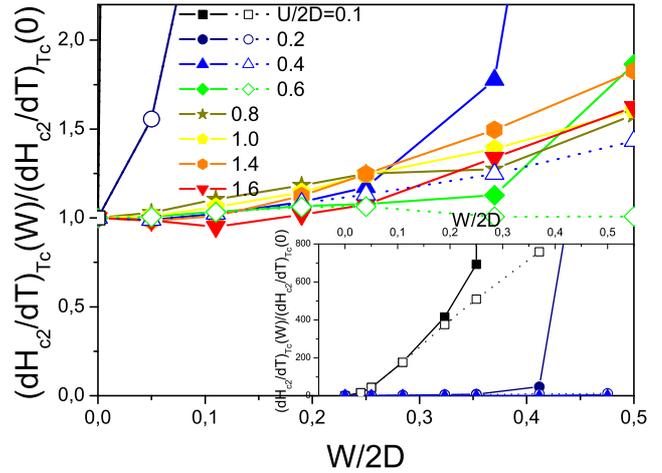}
\caption{Fig. 7. Dependence of the slope of the upper critical field, normalized by
it value in the absence of disorder for different values of Hubbard attraction.
At the insert: the growth of the slope with disorder in the weak coupling limit.}
\label{fig7}
\end{figure}
НIn Fig. \ref{fig7} we show the dependence of the slope of the upper critical
magnetic field on disorder. In the weak coupling limit we again observe
the behavior typical for ``dirty'' superconductors -- the slope of the upper
critical field grows with disorder (cf. insert at Fig. \ref{fig7}).
Taking into account localization corrections in the weak coupling limit greatly
increases the slope of the upper critical field as compares with the ``ladder''
approximation in Anderson insulator ($W/2D \geq$ 0.37). As a result in Anderson
insulator the slope of the upper critical field grows with impurity scattering 
much faster, than in ``ladder'' approximation. At intermediate couplings
($U/2D=$0.4 -- 0.8) the slope of the upper critical field is practically
independent of impurity scattering at weak disorder. In the ``ladder'' approximation
this behavior is conserved also in the region of strong disorder.
However, the account of localization corrections leads to significant growth of the
slope with disorder in Anderson insulator phase. In the limit of very strong coupling
the slope of the upper critical field can even slightly decreases with disorder, but 
for strong disorder the slope grows with the growth of impurity scattering.
In BEC limit the account of localization corrections becomes irrelevant and only
slightly changes the slope of the upper critical field as compared with the
``ladder'' approximation.

\section{Temperature dependence of the orbital upper critical field}

Most vividly the influence of disordering is manifested in the behavior of the 
upper critical field in the theory of ``dirty'' superconductors.
As disorder grows both the slope of the temperature dependence of the upper critical
field at $T_{c}$ \cite{Genn} and $H_{c2}(T)$ at all temperatures increase \cite{WHH,WHH2}. 
Effects of Anderson localization in the limit of strong disorder also are most
explicit in the temperature dependence of the upper critical field.
Precisely at the point of Anderson metal -- insulator transition localization effects
lead to lead to sharp increase of $H_{c2}$ at low temperatures and the temperature
dependence of $H_{c2}(T)$ is qualitatively different from Werthamer, Helfand, Hohenberg
(WHH) dependence \cite{WHH,WHH2}, which is characteristic for the theory of ``dirty'' 
superconductors -- $H_{c2}(T)$ dependence becomes concave \cite{SCLoc_1,SCLoc_2,SCLoc_3}.

Let us consider disorder influence on the temperature dependence of the upper
critical field $H_{c2}(T)$ in a wide region of attraction strength $U$, including
the BCS -- BEC crossover region, as well as for the wide interval of disorders,
up to the vicinity of Anderson transition \cite{JETP17_Hc2}.
In Nozieres -- Schmitt-Rink approach used here the critical temperature of
superconducting transition is determined by a joint solution of equation for
Cooper instability in Cooper particle -- particle channel in weak coupling
approximation and equation for the chemical potential of the system, which is
defined for the whole interval of the values of Hubbard interaction from the
condition of quarter -- filling of the band within DMFT+$\Sigma$ approximation.
The usual condition for Cooper instability has the form:
\begin{equation}
1=-U\chi({\bf q}),
\label{coopinst}
\end{equation}
where $\chi({\bf q})$ is Cooper susceptibility, determined by the loop in Cooper channel.
In the presence of an external magnetic field the total momentum $\bf q$ in Cooper channel
acquires the additional contribution from vector potential $\bf A$ \cite{Genn,WHH}
\begin{equation}
{\bf q}\to{\bf q}-\frac{2e}{c}{\bf A}.
\label{2l}
\end{equation}
As our model assumes an isotropic spectrum, Cooper instability $\chi({\bf q})$ depends
on $\bf q$ only through $q^2$. The minimal eigenvalue of an operator 
${({\bf q}-\frac{2e}{c}{\bf A})}^2$, defining the upper critical magnetic field
$H=H_{c2}$ is \cite{LP}
\begin{equation}
{q_{0}}^2=2\pi\frac{H}{\Phi_{0}},
\label{3l}
\end{equation}
where $\Phi_{0}=\frac{ch}{2e}=\frac{\pi\hbar}{e}$ is magnetic flux quantum. 
Then the equation for $T_{c}(H)$ or $H_{c2}(T)$ remains as usual:
\begin{equation}
1=-U\chi(q^2={q_0}^2).
\label{4l}
\end{equation}

In further analysis we shall neglect the relatively weak influence of magnetic
field on diffusion (noninvariance with respect to time reversal), which is reflected
in nonequality of the loops in Cooper and diffusion channels. This influence of 
magnetic field was analyzed in Refs. работах \cite{SCLoc_3,SCLoc_4,Hc2loc_d3, Hc2loc_d3a}, 
where it was demonstrated, that the account of this, even close to Anderson
metal -- insulator transition, only slightly decreases the value of  $H_{c2}(T)$ in low
temperature region. Under the condition of invariance to time reversal and 
equivalence of the loops in Cooper and diffusion channels, Cooper instability is
determined by the loop in diffusion channel. As a result Eq. (\ref{4l}) for the 
orbital critical field $H_{c2}(T)$ takes the form  \cite{JETP17_Hc2}:
\begin{equation}
1=-\frac{U}{2\pi}\int_{-\infty}^{\infty}d\varepsilon
Im
\left(
\frac{\sum_{\bf {p}}\Delta G_{\bf p}(\varepsilon)}
{2\varepsilon+iD(2\varepsilon)2\pi\frac{H_{c2}}{\Phi_{0}}}
\right)
th\frac{\varepsilon}{2T}.
\label{l5}
\end{equation}
The generalized diffusion coefficient is again determined in the framework
of self -- consistent theory of localization as described above.

\begin{figure}
\includegraphics[clip=true,width=0.48\textwidth]{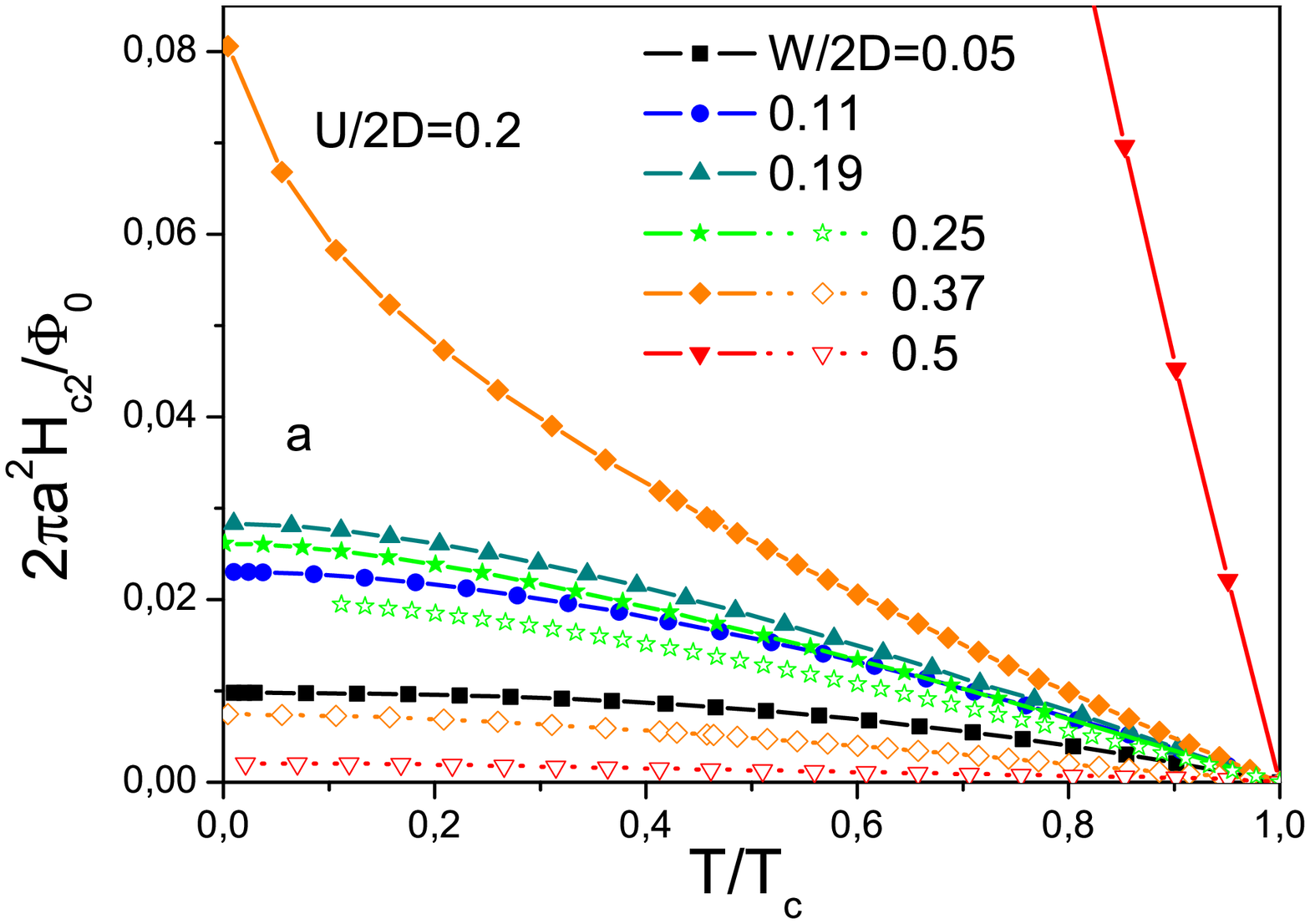}
\includegraphics[clip=true,width=0.48\textwidth]{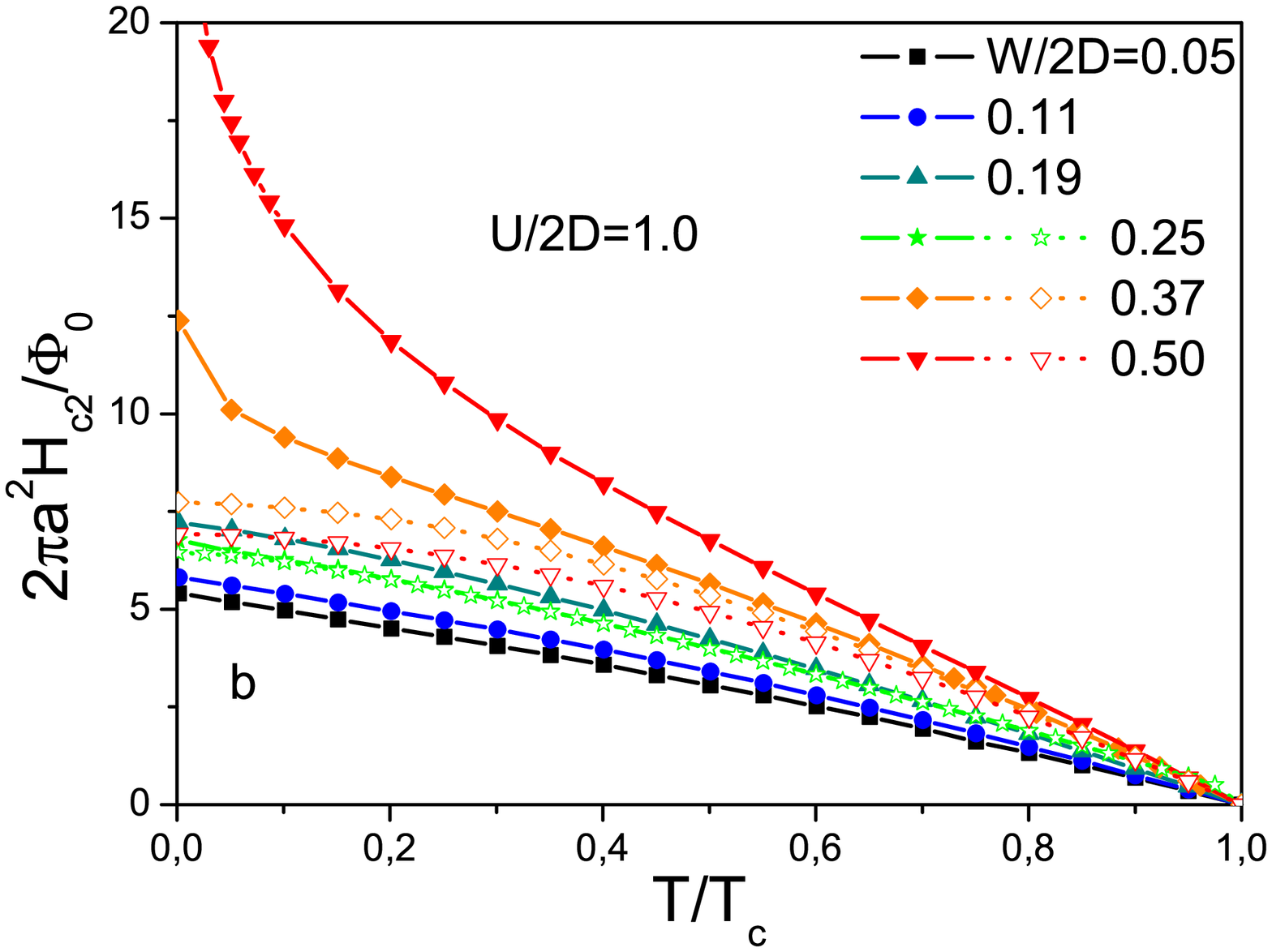}
\includegraphics[clip=true,width=0.48\textwidth]{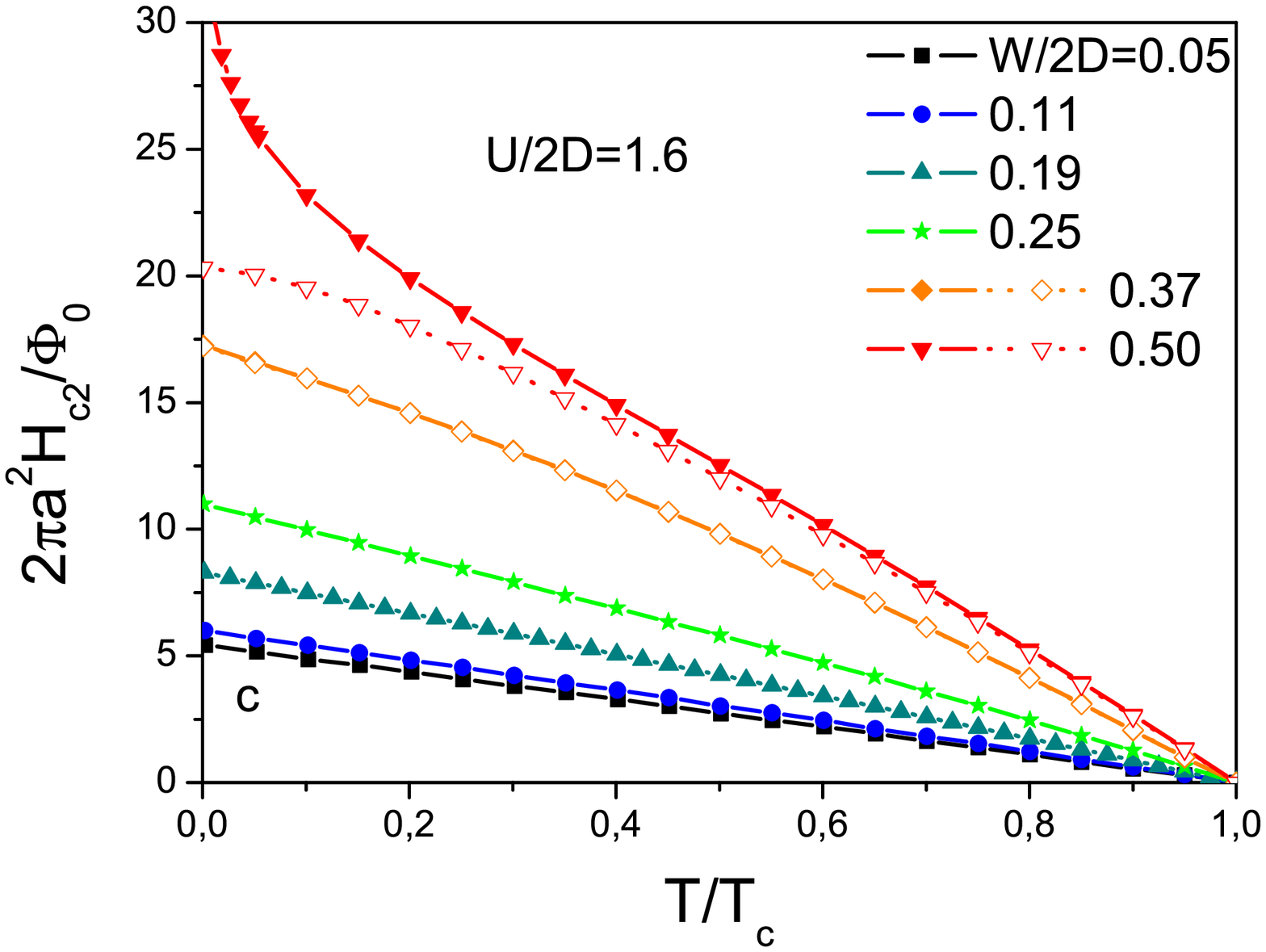}
\caption{Fig. 8. Temperature dependence of the upper critical field
for different values of disorder: 
(a) --  BCS weak coupling limit ($U/2D=0.2$);
(b) --  BCS -- BEC crossover region, intermediate coupling ($U/2D=1.0$);
(c) -- BEC limit of strong coupling ($U/2D=1.6$).
Filled symbols and lines correspond to calculations with account of 
localization corrections. Empty symbols and dashed lines correspond to
``ladder'' approximation for impurity scattering.}
\label{fig8}
\end{figure}
In Fig. \ref{fig8} we show temperature dependencies of the upper critical field
for different degrees of disorder in three regions of coupling strength of
interest to us:
in BCS weak coulping limit ($U/2D=0.2$), in BCS -- BEC crossover region
(intermediate coupling $U/2D=1.0$) and in BEC limit of strong coupling ($U/2D=1.6$).

In strong coupling region (Fig.\ref{fig8}(a)) the growth of disorder leads to
increase of the upper critical field at all temperatures in weak disorder limit
($W/2D<0.19$), in this case the temperature dependencies have negative curvature
and are close in form to the standard WHH dependence \cite{WHH,WHH2}. 
With further growth of disorder and without account of localization corrections
the upper critical field at all temperatures starts to decrease.
However, the account of localization corrections in weak coupling limit at
strong disorder ($W/2D\geq 0.37$) significantly increases the upper critical field
and qualitatively changes its temperature behavior, so that the dependencies.
of $H_{c2}(T)$ acquire the positive curvature. The upper critical field fastly
increases with disorder at all temperatures.

For intermediate coupling (Fig. \ref{fig8}(b)) in the limit of weak disorder
the temperature dependence of the upper critical field becomes practically
linear. The upper critical field at all temperatures increases with the growth
of disorder. In the limit of strong disorder ($W/2D\geq 0.37$) localization
corrections. as in the weak coupling limit, increase the upper critical field
at all temperatures. The dependencies of $H_{c2}(T)$ acquire positive curvature.
However, in the region of intermediate coupling the influence of localization
effects is significantly weaker, than in the limit of weak coupling being
relevant only in the low temperature region.

In BEC limit of strong coupling (Fig. \ref{fig8}(c)) in weak disorder region
$H_{c2}(T)$ dependencies are in fact linear. The upper critical filed grows with
increasing disorder  at all temperatures. In the limit of strong disorder at the
point of Anderson transition itself ($W/2D=0.37$) the dependence $H_{c2}(T)$
remains linear and taking into account localization corrections in fact does not
change the temperature dependence of the upper critical field.
Further increase of disorder leads to the growth of $H_{c2}(T)$. Deep in Anderson
insulator phase ($W/2D=0.5$) $H_{c2}(T)$ dependence acquires positive curvature and
the account of Anderson localization increases $H_{c2}(T)$ in low temperature region,
while close to  $T_c$ localization corrections become irrelevant even at such a
strong disorder. Thus, the strong coupling significantly decreases the influence
of localization effects on the temperature dependence of the upper critical field.

Thus, the increase of coupling strength $U$ leads to a rapid growth of $H_{c2}(T)$, 
especially in low temperature region. In BEC limit and in BEC -- BCS crossover
region $H_{c2}(T)$ dependence becomes practically linear. Disordering at any coupling
strength also leads to the growth of $H_{c2}(T)$. In BCS limit of weak coupling
increasing disorder leads to the growth of both the slope of the upper critical
field close to $T=T_{c}$ and $H_{c2}(T)$ in low temperature region.
In the limit of strong disorder, in the vicinity of Anderson transition
localization corrections lead to the additional sharp increase of the upper
critical field in low temperature region, so that the $H_{c2}(T)$ dependence
becomes concave, acquiring the positive curvature. 
In BCS -- BEC crossover region and in BEC limit weak disorder influence on
the slope of the upper critical field at $T_{c}$ is negligible, though
strong disorder in the vicinity of Anderson transition leads to noticeable
increase of the slope of the upper critical field with disorder.
In low temperature region $H_{c2}(T)$ significantly grows with increasing
disorder, especially in the vicinity of Anderson transition, where localization
corrections noticeably increase $H_{c2}(T=0)$ and $H_{c2}(T)$ dependence becomes
concave, instead of linear, characteristic for the strong coupling at weak disorder.

In the model under discussion the values of the upper critical field at low
temperatures can reach extreme values, up to (or even formally exceeding)
$\frac{\Phi_{0}}{2\pi a^2}$. This requires further analysis of the model, both
taking into account inevitable quantization of electronic spectrum in magnetic
field and paramagnetic effect.

\section{Temperature dependence of paramagnetic critical field}

In weal coupling region and for weak disorder the upper critical magnetic field
of a superconductor is determined by orbital effects and is usually much lower
than paramagnetic limit. However, the growth of the coupling strength and
disorder, as was shown above, lead to a rapid increase of the orbital $H_{c2}$ 
possibly exceeding the paramagnetic limit. In this Section we shall consider
the behavior of paramagnetic critical field for a wide region of coupling strength
$U$, including the region of BCS -- BEC crossover and the limit of very strong
coupling, with the account of disorder (including rather strong one).

It is well known that in BCS weak coupling limit paramagnetic effects (spin splitting
effects) lead to the existence at low temperatures a region on the phase diagram
of a superconductor in magnetic field, where paramagnetic critical magnetic field
$H_{cp}$ decreases with lowering temperature.
This behavior is an evidence of instability, leading to a first order phase transition
to Fulde -- Ferrell -- Larkin -- Ovchinnikov (FFLO) phase \cite{FFLO1,FFLO2,S-Gam_Sarma} 
with Cooper pairs with finite momentum $\bf q$ and periodic in space order parameter.
Further on, our analysis will be limited only to a second order phase transition and 
superconducting order parameter will be assumed spatially homogeneous, allowing us to 
determine the border of instability towards first order transition in the regions of 
BCS -- BEC crossover and strong coupling, also for different levels of disorder.
The problem of stability of FFLO state under these conditions will not be considered.

Within Nozieres -- Schmitt-Rink approach the critical temperature in the presence
of spin splitting in external magnetic field (neglecting orbital effects)
or paramagnetic critical magnetic field $H_{cp}$ at temperature $T<T_c$ is determined
by the following BCS -- like equation \cite{JETP18}:
\begin{eqnarray}
1=\frac{U}{4}\int_{-\infty}^{\infty}d\varepsilon 
\frac{\tilde N_0(\varepsilon)}{\varepsilon -\mu}
\left(
th\frac{\varepsilon -\mu -\mu_BH_{cp}}{2T}+\right. \nonumber\\
\left.+th\frac{\varepsilon -\mu +\mu_BH_{cp}}{2T}
\right),
\label{Hcp}
\end{eqnarray}
where the chemical potential $\mu$ for different values of $U$ and $W$ is determined
from DMFT+$\Sigma$ calculations, i.e. from the standard equation for the number of
electrons in the band. It should be noted that Eq. (\ref{Hcp}) is obtained from an
exact Ward identity \cite{JETP18} and remains valid in the presence of strong disorder,
including the vicinity of Anderson transition. Eq. (\ref{Hcp}) demonstrates, that all
of disorder influence on $H_{cp}$ reduces to renormalization of the bare semielliptic
density of states by disorder, that is for bare band with semielliptic density of
states the influence of disorder on $H_{cp}$ is universal and reduces only to
band widening by disorder, i.e. to the substitution $D\to D_{eff}$.  It is clear that
paramagnetic critical field will be in general rising with the growth of coupling
strength $U$ as it becomes more and more difficult fro magnetic field to break pairs
of strongly coupled electrons \cite{JETP18}.

\begin{figure}
\includegraphics[clip=true,width=0.5\textwidth]{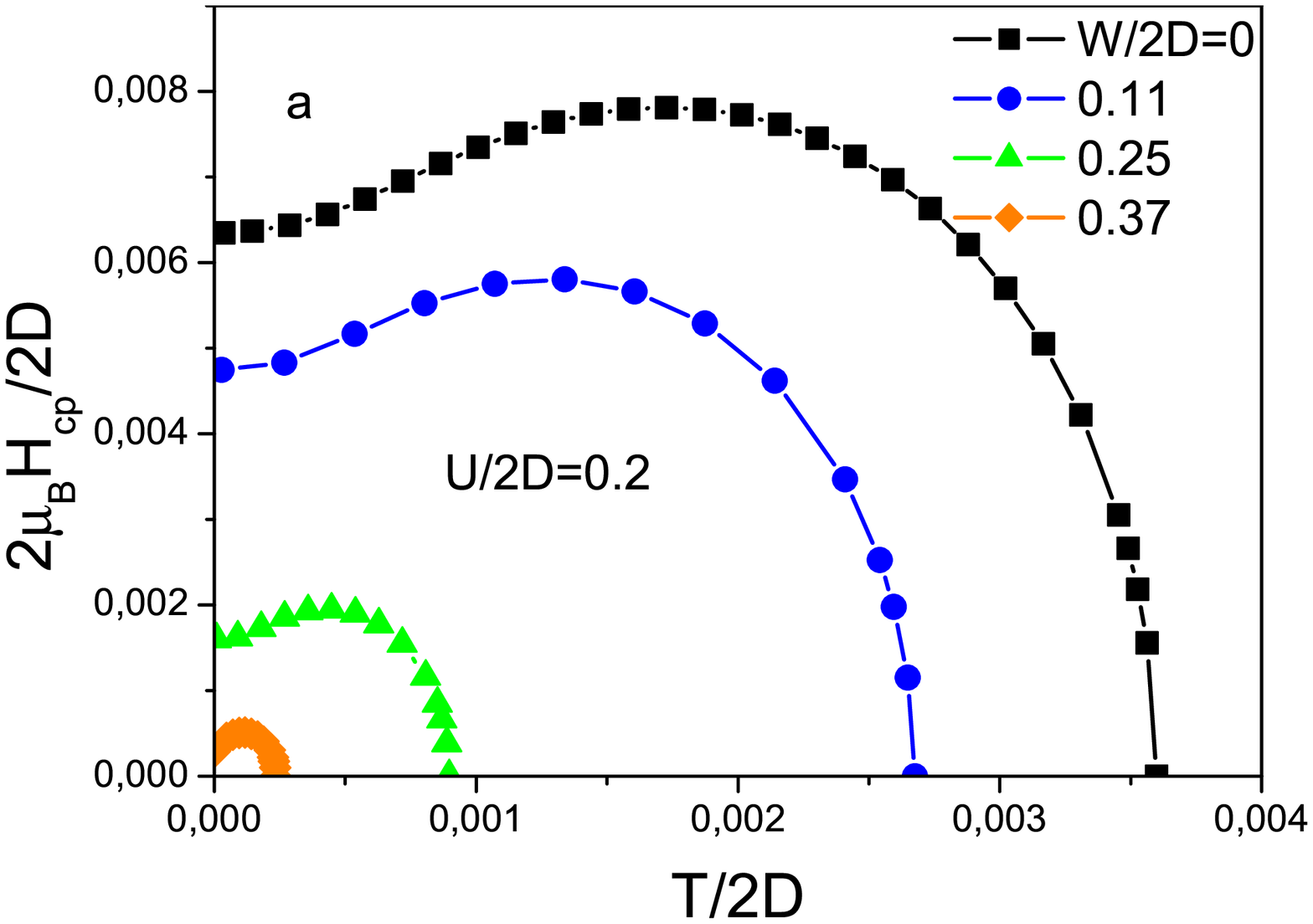}
\includegraphics[clip=true,width=0.48\textwidth]{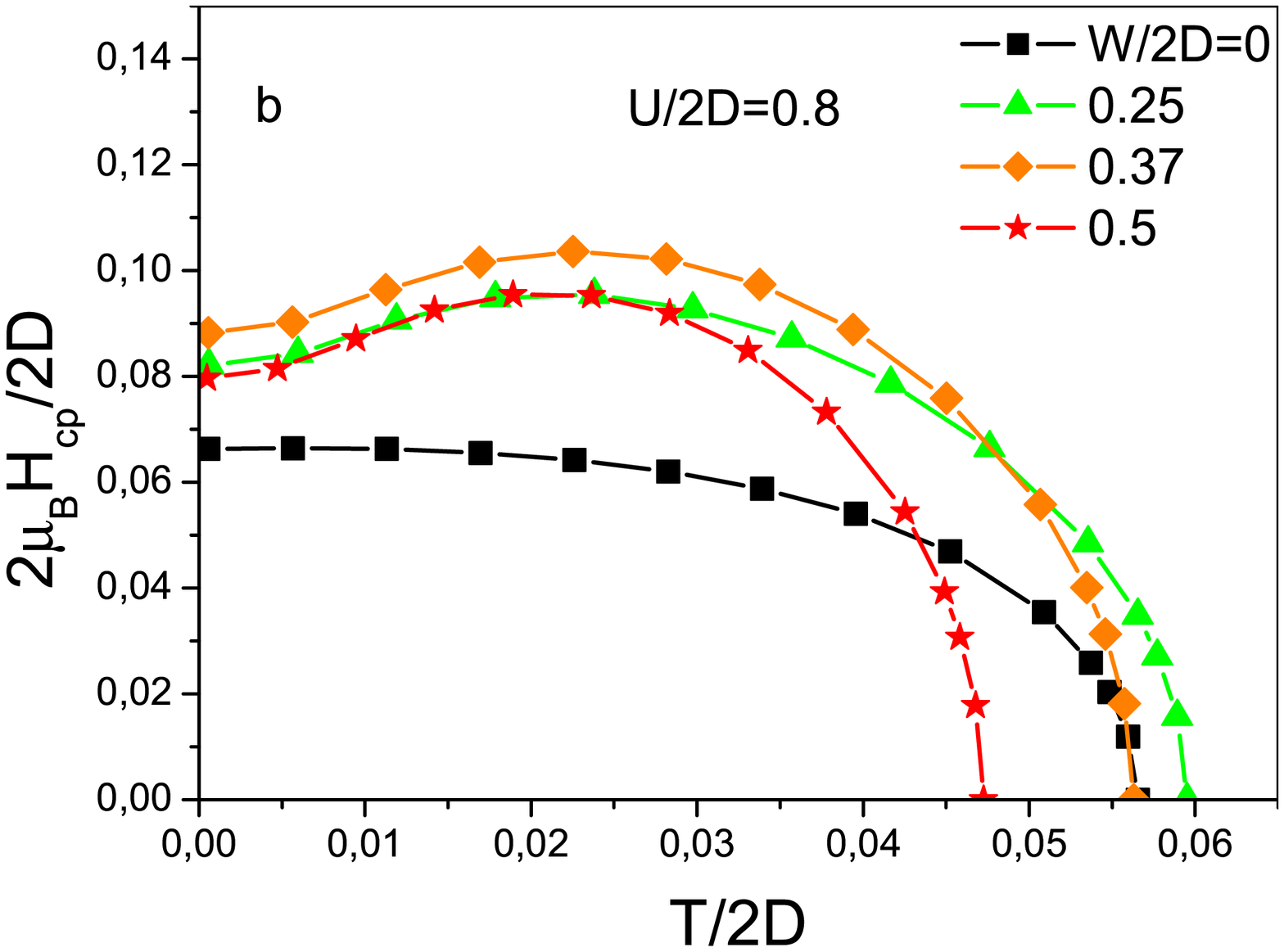}
\includegraphics[clip=true,width=0.48\textwidth]{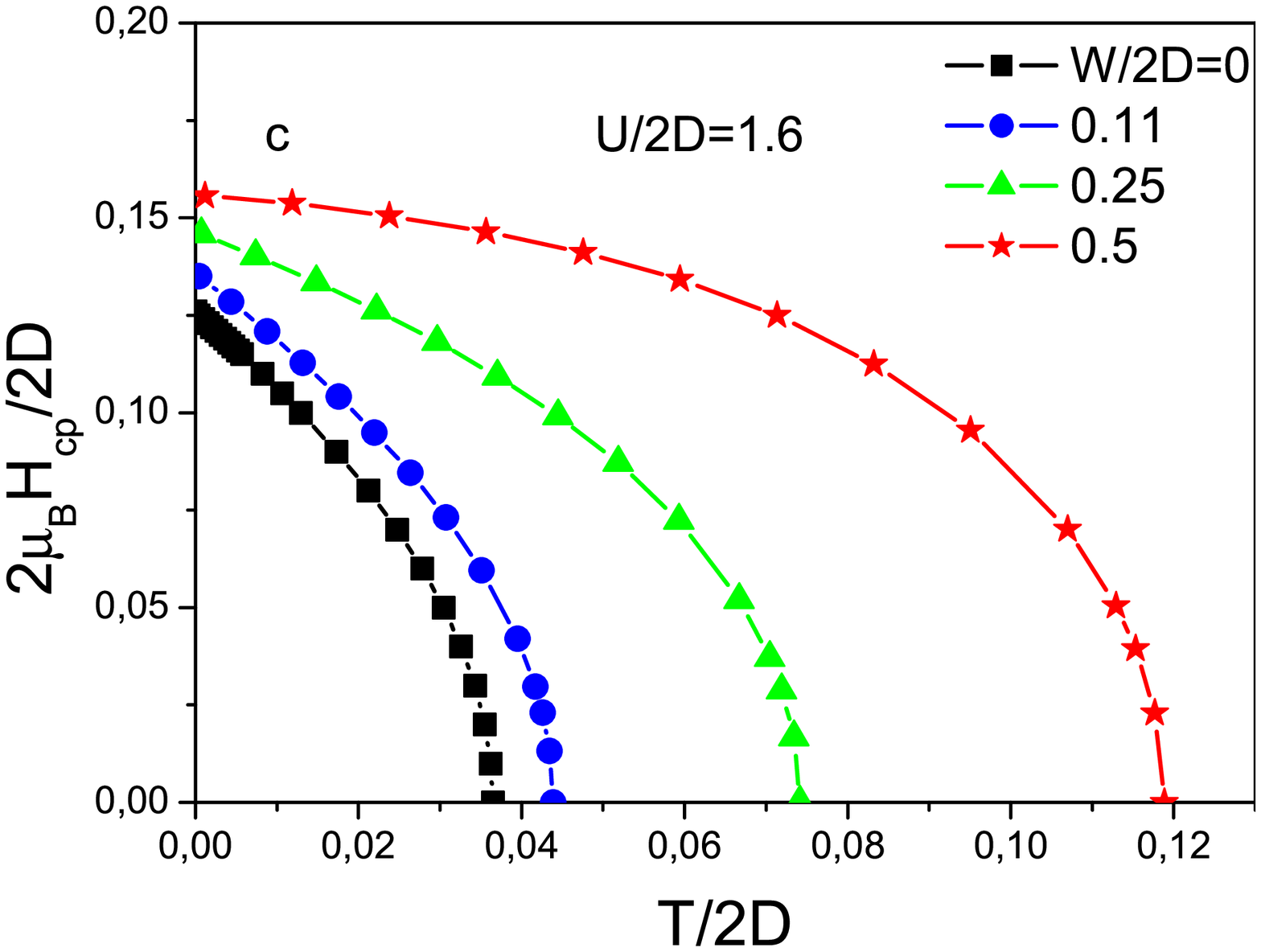}
\caption{Fig. 9. Temperature dependence of paramagnetic critical magnetic field
for different levels of disorder:
(a) --- BCS weak coupling limit ($U/2D=0.2$);
(b) --- BCS -- BEC crossover region (intermediate coupling: $U/2D=0.8$);
(c) --- BEC limit of strong coupling ($U/2D=1.6$).}
\label{fig9}
\end{figure}
In Fig. \ref{fig9} we show the results on disorder influence of temperature
dependence of paramagnetic critical magnetic field.
In BCS weak coupling limit (Fig. \ref{fig9}(a)) disorder growth leads both to
the decrease of critical temperature in the absence of magnetic field $T_{c0}$
(cf. \cite{JTL14,JETP15}) and to the decrease of the critical magnetic field at
all temperatures. Instability region corresponding to first order transition
is conserved also in the presence of disorder. In fact, as was noted before, 
disorder influence upon $H_{cp}(T)$ is universal and related only to the
substitution $D\to D_{eff}$. As a result, the growth of disorder leads to the
decrease of effective coupling strength, which is determined by dimensionless
parameter $U/2D_{eff}$. This leads to a substantial widening of a relative
temperature $T/T_{c}(H)$ region of the first order transition.

For intermediate coupling ($U/2D=0.8$) in the region of BCS -- BEC crossover
(Fig. \ref{fig9}(b)) the growth of disorder only weakly changes the critical
temperature $T_{c0}$ (cf. \cite{JTL14,JETP15}), leading to some increase of
$H_{cp}(T)$. As all influence of disorder is related only to the substitution
$D\to D_{eff}$, the increase of disorder here again leads to the decrease of 
effective coupling strength $U/2D_{eff}$ and restoration of instability region 
of the first order transition.

\begin{figure}
\includegraphics[clip=true,width=0.48\textwidth]{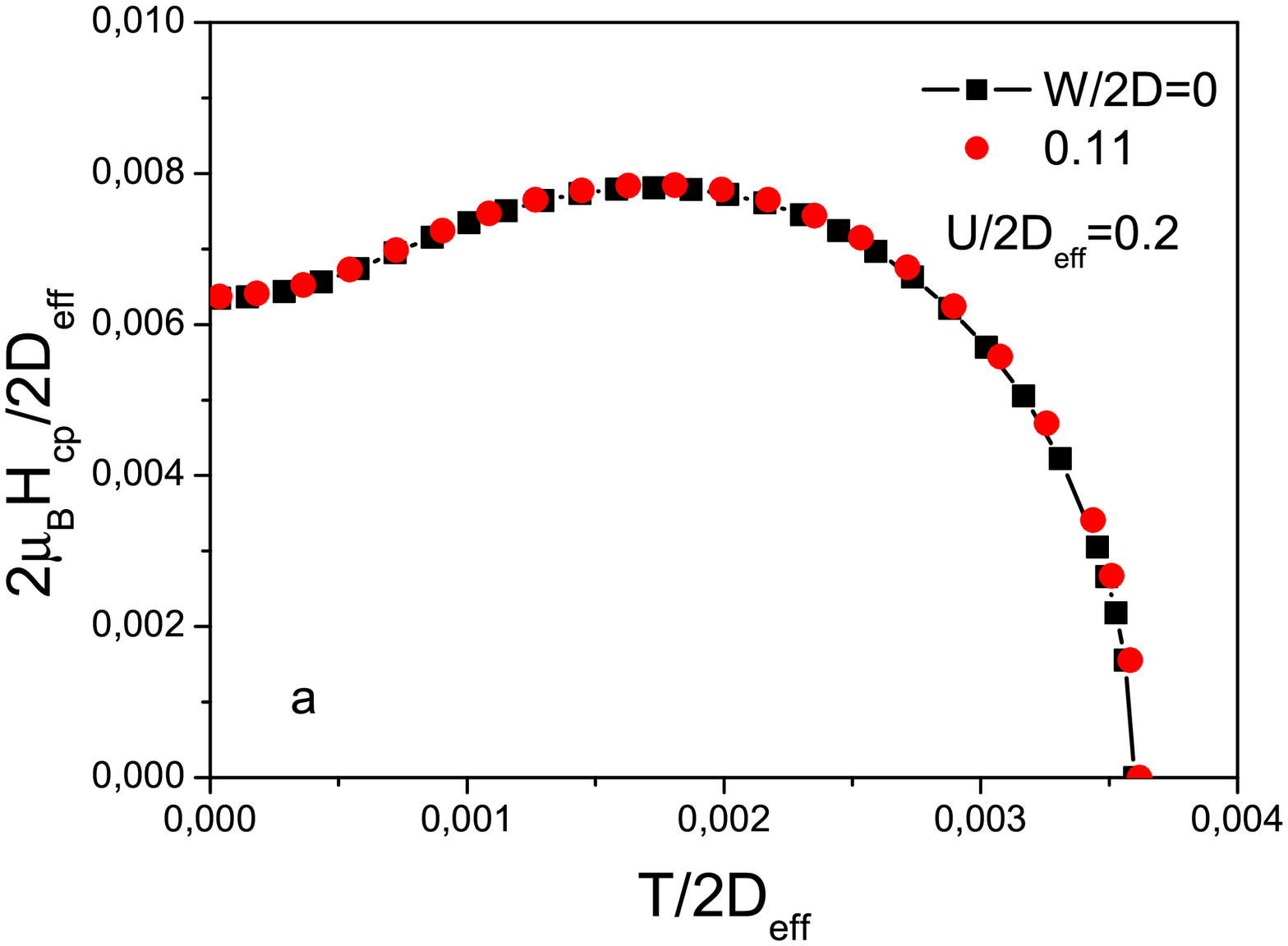}
\includegraphics[clip=true,width=0.48\textwidth]{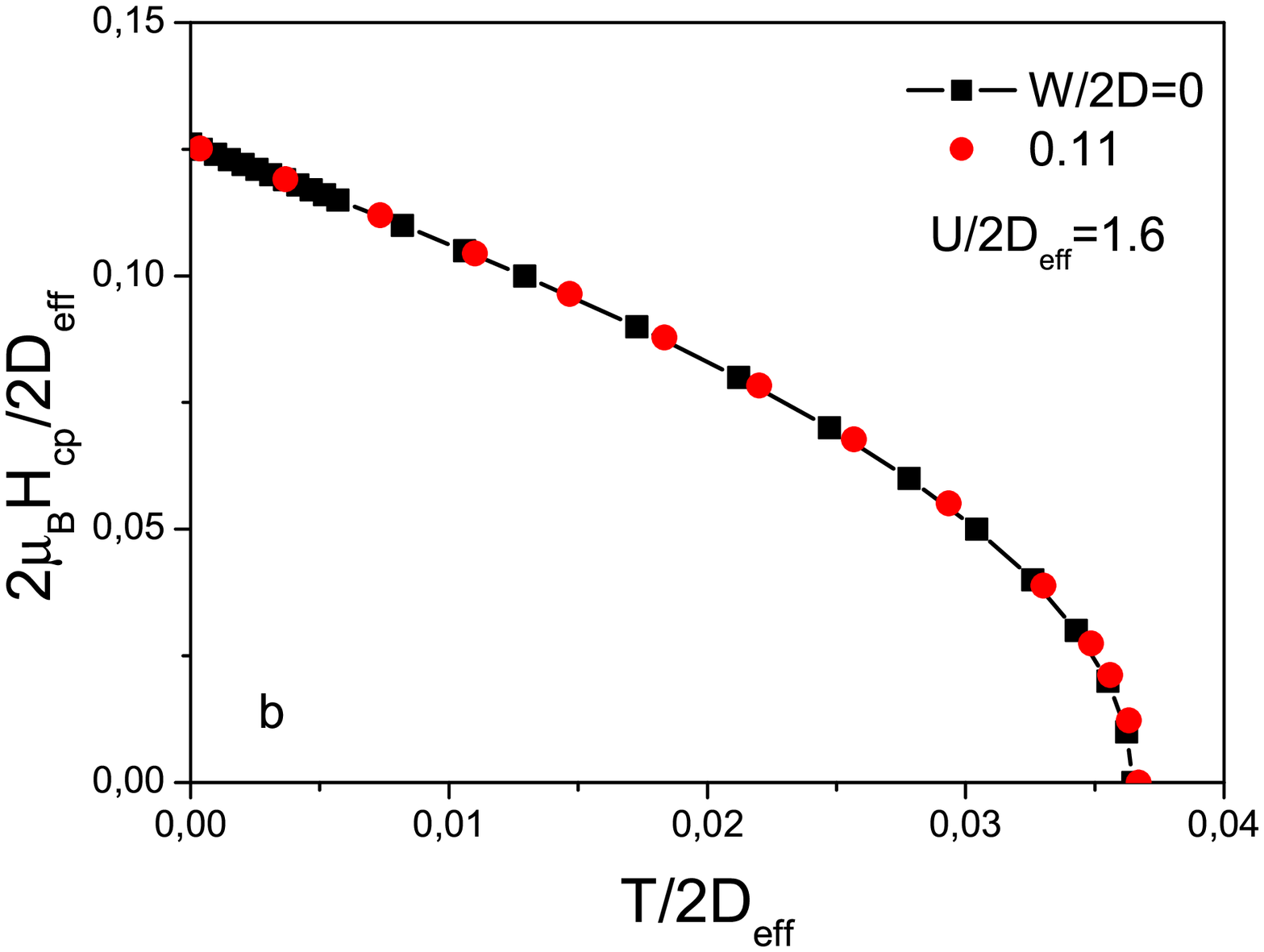}
\caption{Fig. 10. Universal temperature dependence of paramagnetic critical
magnetic field on disorder.
(a) --- weak coupling $U/2D_{eff}=0.2$, $W=0$ and $W=0.11$
(b) --- strong coupling  $U/2D_{eff}=1.6$, $W=0$ and $W=0.11$}
\label{fig10}
\end{figure}

In BEC limit of strong coupling the growth of disorder leads to significant
growth the critical temperature $T_{c0}$ (cf. \cite{JTL14,JETP15}).
At the same time the critical magnetic field in low temperature region is
rather weakly increasing with disorder.
In BEC limit the instability region of first order transition does not appear 
even in the presence of very strong disorder ($W/2D=0.5$). In fact in BEC limit
the influence of disorder is also universal and related only to the substitution
$D\to D_{eff}$. As a result, if we normalize spin splitting and temperature by
effective bandwidth $2D_{eff}$ and fix the effective coupling strength $U/2D_{eff}$, 
we shall obtain the universal temperature dependence of paramagnetic critical
magnetic field. In Fig. \ref{fig10} we show examples of such universal behavior
for typical cases of weak and strong coupling both in presence and in the 
absence of disorder.

In the absence of disorder in BEC limit of strong coupling $U/2D=1.6$ for 
$T\to 0$ we have $2\mu_{B}H_{cp}/2D\approx 0.125$, which for a characteristic
bandwith $2D\sim 1$ eV gives $H_{cp}\sim 10^{7}$ Gauss. For orbital critical
magnetic field (cf. \cite{JETP17_Hc2}) in the same model, for the same coupling
strength and $T\to 0$, for a characteristic value of lattice parameter
$a=3.3*10^{-8}$ cm, we obtain $H_{c2}\approx 1.6*10^{8}$ Gauss.
Thus, the orbital critical magnetic field at low temperatures increases with
coupling strength much faster than paramagnetic field and in BEC  limit the 
main contribution to the upper critical magnetic field at low temperatures 
will be due to paramagnetic effect. The growth of disorder leads to a large
increase of orbital critical magnetic field \cite{JETP17_Hc2}, while 
$H_{cp}(T\to 0)$ in BCS -- BEC crossover region  and in BEC -- limit is
relatively weakly dependent on disorder.
Then, also in the presence of disorder in BEC limit the main contribution
to the upper critical magnetic field at low temperatures will come 
essentially from paramagnetic effect.

Thus, the growth of the coupling strength $U$ leads to rapid increase of 
$H_{cp}(T)$ and disappearance of the instability region of first order transition
at low temperatures in BCS -- BEC crossover region and in BEC limit, which appears
at low temperatures in BCS limit of weak coupling. Physically this is related to the
fact, that it is more difficult for magnetic field to break strongly coupled pairs.
The growth of disorder on BCS limit of weak coupling leads both to the decrease of
critical temperature and to the decrease of $H_{cp}(T)$.
Instability region of first order transition at low temperatures in the presence
of disorder is conserved. In the region of intermediate coupling ($U/2D=0.8$) 
disorder influence on both critical temperature and $H_{cp}(T)$ is rather weak.
However, the growth of disorder leads to restoration of low temperature region of
instability of the first order transition, which is not observed in the absence of
disorder. This rather unexpected conclusion is due to specifics of attractive 
Hubbard model, where the effective dimensionless parameter $U/2D_{eff}$ controls
the coupling strength in disordered case.

In BEC limit at low temperatures, for reasonable parameters of the model, 
paramagnetic critical magnetic field is noticeably lower than the orbital one,
so that the upper critical field in this region is determined essentially by
paramagnetic critical field. In the presence of disorder this conclusion is
also even more valid, as the orbital critical field rapidly grows with disorder,
while paramagnetic critical field in this limit only weakly dependent on disorder.

\section{Conclusion}

In this paper, within Nozieres -- Schmitt-Rink approximation and DMFT+$\Sigma$ 
generalization of dynamic mean field we have studied the influence of
disordering, including the strong one (Anderson localization region), on 
Ginzburg -- Landau expansion and the behavior of related physical properties
close to $T_c$, and also the upper critical magnetic field (both orbital and
paramagnetic) in disordered Anderson -- Hubbard model with attraction,
for a wide range of the values of attraction potential $U$, from the region of
weak coupling, where instability of the normal phase and superconductivity are well
described by BCS model, up to the limit of strong coupling, where superconducting 
transition is related to Bose condensation of compact Cooper pairs, which are
formed at temperatures much higher than superconducting transition temperature.

Due to size limitations of this review above we have presented only a part of 
our results. Further details, as well as more detailed derivations of the main
equations can be found in original papers
\cite{JETP16,FNT16,JETP17,JETP17_Hc2,JETP18}.

Note that all results obtained in this work implicitly used an assumption
of self -- averaging superconducting order parameter entering Ginzburg -- Landau
expansion. It is well known \cite{SCLoc_3} that this assumption becomes, in general
case, invalid close to Anderson metal -- insulator transition, which is due to
development in this region of strong fluctuations of the local density of
states, leading to strong spatial fluctuations of the order parameter \cite{NAV_1} 
and inhomogeneous picture of superconducting transition \cite{NAV_2}.
This problem is of great interest in the context of superconductivity in
BCS -- BEC crossover and in the region of strong coupling, and deserves further
studies.

This work was supported in part by RFBR grant 20-02-00011.

\newpage

\end{document}